\documentclass[sigconf]{acmart}
\usepackage{multirow}
\usepackage{booktabs}
\usepackage{colortbl}
\usepackage{enumitem}
\usepackage{soul} 
\usepackage{color} 
\usepackage{xcolor}
\definecolor{bestresult}{RGB}{221,238,211}
\definecolor{Green}{RGB}{112,173,71}
\definecolor{Red}{RGB}{237,125,49}
\definecolor{Purple}{RGB}{112,48,160}
\definecolor{Blue}{RGB}{101,165,223}
\definecolor{DarkBlue}{RGB}{68,114,196}
\definecolor{Yellow}{RGB}{255,192,0}
\usepackage{listings}
\acmSubmissionID{2459}

\setcopyright{none}
\renewcommand\footnotetextcopyrightpermission[1]{} 
\begin{document}

\title{Visual Captioning at Will: Describing Images and Videos \\Guided by a Few Stylized Sentences}


\author{Dingyi Yang}
\authornote{This work was completed during the author's internship at Alibaba Group.}
\email{yangdingyi@ruc.edu.cn}
\affiliation{%
  \institution{School of Information,\\ Renmin University of China}
  \state{Beijing}
  \country{China}
  \postcode{43017-6221}
}

\author{Hongyu Chen}
\email{yinchen.chy@alibaba-inc.com}
\affiliation{%
  \institution{Alibaba Group}
  \state{Beijing}
  \country{China}
}

\author{Xinglin Hou}
\email{ xingli.hxl@alibaba-inc.com}
\affiliation{%
  \institution{Alibaba Group}
  \state{Beijing}
  \country{China}
}
\author{Tiezheng Ge}
\email{ tiezheng.gtz@alibaba-inc.com}
\affiliation{%
  \institution{Alibaba Group}
  \state{Beijing}
  \country{China}
}
\author{Yuning Jiang}
\email{ mengzhu.jyn@alibaba-inc.com}
\affiliation{%
  \institution{Alibaba Group}
  \state{Beijing}
  \country{China}
}

\author{Qin Jin}
\authornote{Corresponding Author.}
\email{qjin@ruc.edu.cn}
\affiliation{%
  \institution{School of Information, \\Renmin University of China}
  \state{Beijing}
  \country{China}
  \postcode{43017-6221}
}


\begin{abstract}
  Stylized visual captioning aims to generate image or video descriptions with specific styles, making them more attractive and emotionally appropriate. One major challenge with this task is the lack of paired stylized captions for visual content, so most existing works focus on unsupervised methods that do not rely on parallel datasets. However, these approaches still require training with sufficient examples that have style labels, and the generated captions are limited to predefined styles. To address these limitations, we explore the problem of Few-Shot Stylized Visual Captioning, which aims to generate captions in any desired style, using only a few examples as guidance during inference, without requiring further training. We propose a framework called FS-StyleCap for this task, which utilizes a conditional encoder-decoder language model and a visual projection module. Our two-step training scheme proceeds as follows: first, we train a style extractor to generate style representations on an unlabeled text-only corpus. Then, we freeze the extractor and enable our decoder to generate stylized descriptions based on the extracted style vector and projected visual content vectors. During inference, our model can generate desired stylized captions by deriving the style representation from user-supplied examples. Our automatic evaluation results for few-shot sentimental visual captioning outperform state-of-the-art approaches and are comparable to models that are fully trained on labeled style corpora. Human evaluations further confirm our model’s ability to handle multiple styles.

\end{abstract}



\settopmatter{printfolios=true}
\maketitle

\section{Introduction}
Traditional visual captioning \cite{guo2020normalized,hu2022scaling,wang2022git} aims to describe visual content with objective and neutral explanations, denoted as factual captions. 
However, recent research \cite{gan2017stylenet,mathews2016senticap,bin2021multi} has demonstrated that incorporating style into visual captioning can greatly enhance the user experience. By adding style, the captions become more attractive and emotionally appropriate. For example, ``The broccoli added with the chicken makes this the best pizza!'' is more appealing than ``A pan of pizza with broccoli and cheese''. Therefore, Stylized Visual Captioning ~\cite{mathews2016senticap,gan2017stylenet,wu2023sentimental} has attracted increasing research interest.

\begin{figure}[t]
  \centering
  \includegraphics[width=\linewidth]{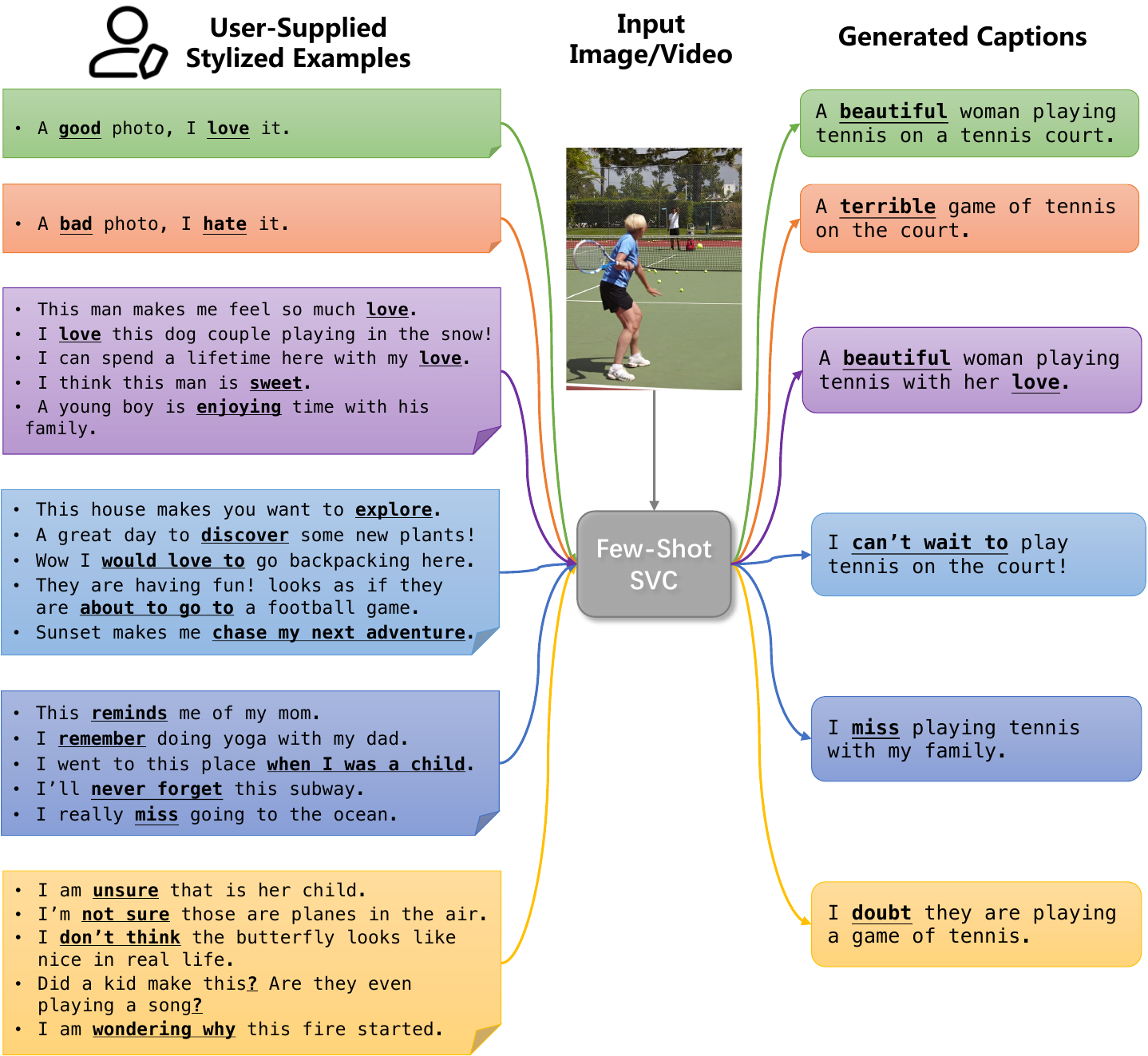}
  \vspace{-8pt}
  \caption{Few-Shot Stylized Visual Captioning accepts a small number of stylized examples from users, and generates descriptions for images or videos with the desired style, similar to the input samples. We present examples with different styles generated by our proposed FS-StyleCap, including {\color{Green}{Positive}}, {\color{Red}{Negative}}, {\color{Purple}{Romantic}}, {\color{Blue}{Adventurous}}, {\color{DarkBlue}{Memorable}}, and {\color{Yellow}{Skeptical}} styles from top to bottom.}
  \Description{Task}
  \label{fig:task}
\end{figure}

Previous stylized visual captioning works can be divided into two categories: supervised and unsupervised methods. Supervised methods use a parallel training set to learn stylized descriptions for visual inputs. These methods are limited in the range of style types covered by existing datasets \cite{mathews2016senticap, gan2017stylenet}. As it is time-consuming and expensive to manually compose paired stylized captions for large-scale images or videos, most works have shifted toward unsupervised learning \cite{mathews2018semstyle,tan2022detach}, relying on an unpaired style-specific training corpus and a factual vision-caption dataset. Although these methods eliminate the need for paired stylized data, they are restrained by the pre-specified style labels of the stylized-text corpora. In other words, to generate descriptions with a new desired style, the models need to be trained or fine-tuned on a corresponding stylized-text corpus.

In this work, we focus on a more practical formulation of stylized visual captioning, namely Few-Shot Stylized Visual Captioning (\textbf{Few-Shot SVC}). As illustrated in Figure \ref{fig:task}, Few-Shot SVC aims to describe images or videos in any target style specified by a few examples during inference time. Our method directly extracts the text style from user-supplied examples and guides caption generation to meet the desired style. It even eliminates the need for training on a curated text-only corpus with annotated style attributes. Although more challenging than previous works, Few-Shot SVC provides significantly more flexibility.

\begin{figure*}[h]
  \centering
  \includegraphics[width=0.9\textwidth]{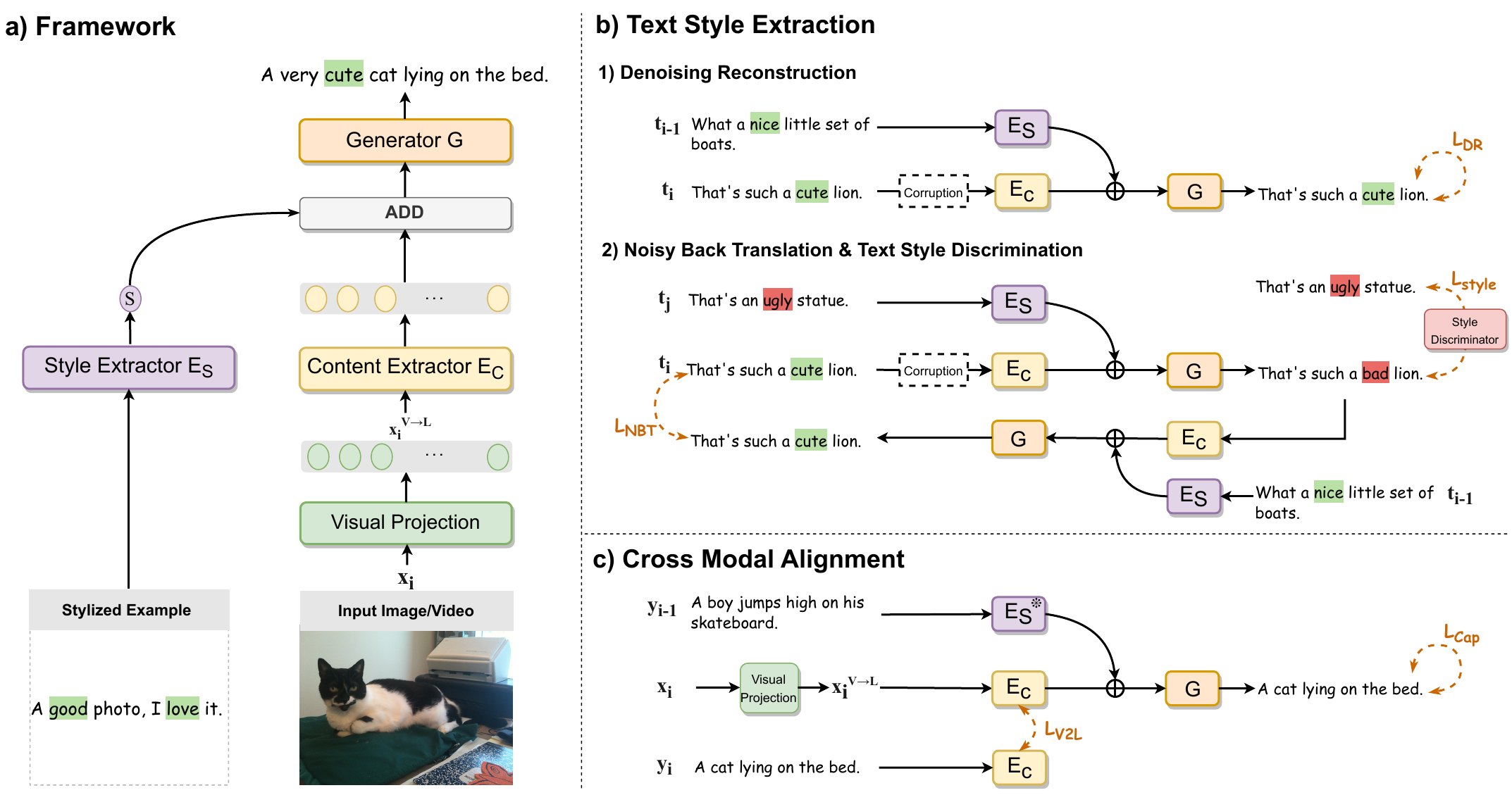}
  \vspace{-10pt}
  \caption{Overview of our proposed FS-StyleCap, which is trained on an unlabled text corpus and a factual vision-caption datatset. a) The overall framework consists of four key modules: the style extractor $E_s$, content extractor $E_c$, generator $G$ and visual projection module $M^{V \rightarrow L}$; 
  b) In the first training stage, we aim to train $E_s$ using the text-only corpus. We include three tasks (in Sec \ref{sec:style_extraction}): denoising reconstruction, noisy back translation and text style discrimination;
  c) In the second training stage, we freeze $E_s$, and train other modules with factual visual captioning tasks (in Sec \ref{sec:visual_task}) to achieve cross-modal alignment. To ensure the ability to generate stylized descriptions, we simultaneously train $E_c$ and $G$ with text style injection tasks as illustrated in b), while also keeping $E_s$ frozen.}
  \Description{Framework}
  \label{fig:frame}
\end{figure*}

Few-shot SVC faces two main challenges: (1) extracting style information from a small amount of example sentences, and (2) generating stylized descriptions that are compatible with both style information (textual modality) and visual content information (visual modality).
To address these challenges, we propose a framework called \textbf{FS-StyleCap}, which is trained on a text-only corpus without style labels, as well as a factual vision-caption dataset. This framework utilizes a conditional encoder-decoder language model that extracts style representations to guide the generation of stylized captions. Additionally, it employs a visual projection module to enable cross-modal alignment. We apply a two-step training scheme. First, we use the unlabeled text-only corpus to train a style extractor that can extract style representations from textual inputs. The extractor is fine-tuned from the encoder of a large-scale language model \cite{raffel2020exploring} with a strong ability to represent  text.
Inspired by TextSETTR~\cite{riley2020textsettr}, we assume that two adjacent sentences from the unlabeled corpus possess the same style. We train our model to extract a style vector from the first sentence and use it to perform reconstruction tasks on the second sentence. To further improve the quality of the style vector, we introduce a style-related loss, computed by a style discriminator that has been trained using contrastive learning \cite{hadsell2006dimensionality}. In the second stage, we freeze the style extractor, and address the problem of cross-modal alignment. As illustrated in Figure \ref{fig:frame}, we project high-level visual features into the joint embedding space, and then extract visual content vectors. The caption generator fuses the visual content vectors and a text style vector to generate a stylized description. Multi-task training strategy is applied to ensure the capacity for style injection and modality alignment. In detail, we simultaneously train on visual captioning tasks using the factual vision-caption dataset, and on text style injection tasks using the unlabeld style corpus. During inference, we extract the desired style vector from a few examples and use it to guide the generation of stylized visual captions.

In summary, the contributions of this work are as follows:
\parskip=0.1em
\begin{itemize}[itemsep=0.2em,parsep=0em,topsep=0em,partopsep=0em,leftmargin=1em,itemindent=0.2em]
\item To the best of our knowledge, this is the first work to tackle Few-Shot Stylized Visual Captioning. This task uses a few example sentences to guide the generation of stylized captions, without requiring further training.
\item We propose a framework called FS-StyleCap, which extracts the style representation from text input and aligns visual information with the text style vector to generate stylized captions.
\item The automatic evaluation results of sentimental visual captioning outperform state-of-the-art models with few-shot stylized examples. Our approach even achieves comparable results with models that are fully trained on labeled style corpora.
\item As verified by human evaluation, our method demonstrates strong performance on multiple style attributes, with only 1-5 example sentences as guidance. 
\end{itemize}
\section{Related Works}

\noindent \textbf{Stylized Visual Captioning.} 
Stylized Visual Captioning \citep{mathews2016senticap,bin2021multi} aims to generate descriptions of images or videos that are both visually relevant and stylistically accurate. Previous works in this area can be categorized into supervised and unsupervised methods. Supervised approaches \cite{mathews2016senticap,you2018image,shuster2019engaging,li2021similar,bin2021multi,li2022taking} involve constructing paired stylized captions for visual content. However, building large-scale paired datasets can be laborious, and these methods are limited to certain styles in parallel datasets.

Therefore, most existing works focus on unsupervised methods. \citet{mathews2018semstyle} propose to generate semantic terms for visual content, and transform them into descriptions with various styles. Several approaches, including StyleNet \cite{gan2017stylenet}, Factual \cite{chen2018factual}, DLN \cite{chen2019unsupervised}, and MSCap \cite{guo2019mscap}, suggest different architectures for learning style-dependent matrices that capture style-related information.
MemCap \cite{zhao2020memcap} and Senti-Trans \cite{wu2023sentimental} incorporate style knowledge to generate stylized descriptions.
\citet{tan2022detach} propose to detach several text style representations for specific styles, and then attach them to image content to generate stylized captions. Recently, some works start to leverage the power of large language models. \citet{zeng2023conzic} propose a sampling-based methods to generate captions that incorporates controllable signals like predefined sentiments. \citet{nukrai2022capdec} and \citet{gu2022can} utilize the CLIP embedding space \cite{radford2021clip} and additional stylized training data to perform stylized captioning tasks, achieving impressive results. Although these unsupervised methods have shown effective results, they still require sufficient samples with desired style labels, and additional training is necessary to accommodate new styles. These limitations motivate us to explore the problem of few-shot stylized visual captioning. 

\noindent \textbf{Few-shot Text Style Transfer.} 
Recently, some works have started to explore the problem of few-shot text style transfer. This task does not require style labels during training, but instead uses a small number of labeled stylized examples as guidance during inference.
\citet{xu2020variational} propose to train a variational auto-encoder on unlabeled text, learning a text representation that features a controllable portion, which is restricted to lie on a k-dimensional simplex. To perform transfer, the controllable dimensions are manipulated with a basis vector that corresponds most strongly to the target style. 
Similarly, \citet{riley2020textsettr} fine-tune the T5 encoder \cite{raffel2020exploring} to extract a style vector from input text. This vector can then be employed to modify the latent representation in order to generate text with the target style. 
\citet{reif2021recipe}, \citet{luo2023prompt} and \citet{suzgun2022prompt} employ prompting-based methods. They utilize large language models (LLMs) and design specific prompts to rewrite texts in various styles. However, these methods usually require paired prompts that represent the relations and differences between different styles, which makes the process less flexible. Additionally, they face the problem of hallucinations, making them less reliable \cite{reif2021recipe} when compared to trained methods. To address these concerns, we utilize pre-trained models and fine-tune them to tackle our specific task, leveraging the benefits of LLMs and avoiding the drawbacks of prompt-based techniques at the same time.

\vspace{-2pt}
\section{Method}

Given the visual content (an image or a video) $x$, our goal is to generate a stylized description in the target style specified by a few example sentences. 
As the (vision, stylized-caption) paired data is not available for training, we leverage a factual captioning dataset that consists of  (vision, factual-caption) paired data denoted as $\mathcal{V}=\{(x_i, y_i)\}_{i=1}^N$, 
as well as a text-only corpus that contains miscellaneous text without style labels, denoted as $\mathcal{T}=\{t_i\}_{j=1}^{M}$ . 

\vspace{-6pt}

\paragraph{\textbf{Overall Framework}}\label{sec:framework} 
In Figure \ref{fig:frame} (a), we illustrate the framework of our proposed FS-StyleCap. It comprises four key modules: (1) the \textit{Style Extractor $E_s$} that extracts the style-related information from text and represents it as a single style vector; 
(2) the \textit{Content Extractor} $E_c$ that extracts the content-related vectors; (3) the \textit{Generator} $G$, which accepts the input by fusing \footnote{In detail, we mean-pool the output hidden states of $E_s$ to obtain a style vector $s$, and add it to each content vector.} the content vectors and the style vector to generate captions in the target style; (4) a \textit{Visual Projection Module} $M^{V \rightarrow L}$ that projects the visual features to the shared visual-language embedding space. $E_c$/$E_s$ and $G$ can all leverage the power of large-scale language models, i.e. initialized with the weights of pre-trained T5 \cite{raffel2020exploring}. Specifically, $E_c$ and $E_s$ are initialized with the pre-trained encoder, while their abilities are differentiated after our fine-tuning; $G$ is initialized with the decoder, and can generate stylized descriptions conditioned on an additional style vector after training.

Due to the lack of (vision, stylized-caption) paired data and style-labeled text for training, we leverage the available factual caption dataset $\mathcal{V}$ and a textual corpus $\mathcal{T}$ without style labels.   
To fully train the core components $E_s$, $E_c$, and $G$, we propose a two-stage training scheme for our framework. The style extractor $E_s$ only requires knowledge from the text modality. Thus in the first training stage (Section \ref{sec:style_extraction}), we use an unlabeled style corpus to train $E_s$. This stage involves three tasks: denoising reconstruction, noisy back translation, and style discrimination. In the second stage (Section \ref{sec:cross_modality}), we freeze $E_s$ and train other modules simultaneously with Visual Captioning tasks on $\mathcal{V}$ and Text Style Injection tasks on $\mathcal{T}$. As a result, our model can generate descriptions in any style based on the extracted style vector and visual content vectors.
\subsection{Text Style Extraction}\label{sec:style_extraction}
Large language models possess a strong ability in text representation \cite{riley2020textsettr,suzgun2022prompt}. Our objective is to distill the encoder's ability to extract style-related representations. 
To achieve this with an unlabeled corpus, we assume two adjacent sentences $t_i$ and $t_{i-1}$ in the same document possess the same style \cite{riley2020textsettr}. This allows us to perform reconstruction tasks for $t_i$ based on the style vector of $t_{i-1}$. Note that the text corpus also includes factual captions, all of which possess the factual style \cite{gan2017stylenet,tan2022detach}. This enables our style extractor to handle factual style as well. We train our text style extractor $E_s$ through three tasks as follows.
\paragraph{\textbf{Denoising Reconstruction}} Each sentence $t_i$ is decomposed into a sequence of discrete tokens $\{w^i_{1}, \ldots, w^i_{k_i}\}$ by the pre-trained tokenizer. Our noise function first corrupts the input sentence by dropping tokens in $t_i$ with probability $p$. Then, the content extractor $E_c$ extracts content-related vectors from the corrupted sentence $\tilde{t}_i$. As sentence $t_{i-1}$ is assumed to have the same style as $t_i$, the stylized generator attempts to reconstruct $t_i$ conditioned on the style vector of $t_{i-1}$. Since noise applied to the input $t_i$ can corrupt words that convey the input style, the decoder must learn to use the additional style vector to generate stylized descriptions. This will also encourage $E_s$ to extract style-related information. Our goal is to minimize the cross-entropy loss as follows:
\begin{equation}
\begin{split}  
\mathcal{L}_{DR} &= \mathcal{L}_\textrm{CE}\bigg(t_i, G\Big(E_c\big(\tilde{t}_i\big),E_s\big(t_{i-1}\big)\Big)\bigg) \\
&= -\sum_{n=1}^{K_i} \textrm{log } p_{\theta_G, \theta_{E_c}, \theta_{E_s}}(w_n^i|w_{1:n-1}^i,\tilde{t}_i,t_{i-1}),
\end{split}
\end{equation}
where $\theta_G, \theta_{E_c}, \theta_{E_s}$ are the parameters of generator $G$, content extractor $E_c$, and style extractor $E_s$. 

\paragraph{\textbf{Noisy Back Translation}} First, we corrupt $t_i$ to obtain $\tilde{t}_i$. Then, we apply the current model to transfer $\tilde{t}_i$ into $t'_i$ with another text style, using the style vector of a randomly sampled $t_j$. Finally, we apply our model to translate $t'_i$ back into the original style, as instructed by the style vector of $t_{i-1}$. As verified by \citet{riley2020textsettr}, this approach encourages the model to identify which tokens in the input $t'_i$ do not match the target style indicated by $t_{i-1}$ and change them. This improves the effectiveness of the style extractor. The loss function is defined as follows:
\begin{equation}
\begin{split}
\mathcal{L}_{NBT} &= \mathcal{L}_\textrm{CE}\bigg(t_i, G\Big(E_c\big(t'_i\big),E_s\big(t_{i-1}\big)\Big)\bigg)\\
 &= -\sum_{n=1}^{K_i} \textrm{log } p_{\theta_G, \theta_{E_c}, \theta_{E_s}}(w_n^i|w_{1:n-1}^i,t'_i,t_{i-1}) .  
\end{split}
\end{equation}

\paragraph{\textbf{Text Style Discrimination}} To further improve the quality of extracted style vectors, we propose a style-related loss. To achieve this, we train a style discriminator $D$, which can compute the similarity in text style between two sentences.  During the process of noisy back translation, $t'_i$ should have the same style as $t_j$ as it is generated based on the style vector of $t_j$. 
We calculate the style discrimination loss as follows:
\begin{equation}
    \mathcal{L}_{style} = -logD(t'_i, t_j),
\end{equation}
where the discriminator\footnote{The discriminator is fine-tuned with the RoBERTa-base \cite{liu2019roberta} model, applying the embedding of [CLS] token as sentence embedding.} is trained using contrastive learning \cite{hadsell2006dimensionality}, with positive samples from the same paragraph, and negative samples from other sentences in the training batch.
\paragraph{\textbf{Final Loss}}
The final loss term used for training is the average of the three losses, denoted as:
\begin{equation}\label{eqn:l_text}
   \mathcal{L}_{Text} =( \mathcal{L}_{DR}+ \mathcal{L}_{NBT}+\mathcal{L}_{style})/3.
\end{equation}
\subsection{Cross-Modal Alignment}\label{sec:cross_modality}
Once we have the style extractor, which can obtain the style representation of any input text, the main challenge is to enable the generator to produce stylized captions based on the  extracted style vector and visual inputs. We use visual feature projection to achieve cross-modal alignment, and propose a multi-task training strategy to ensure the ability of style injection simultaneously. During this training process, the style extractor $E_s$ is frozen, while the content extractor $E_c$ and caption generator $G$ are fine-tuned. 
\paragraph{\textbf{Visual Feature Projection}} \label{sec:visual_task} A transformer-based module $M^{V \rightarrow L}$ projects the visual representation to the appropriate input dimension for our language model. Concretely, we feed the transformer network with a concatenation of two inputs: the visual representation and a learnable constant embedding. The constant vector will extract semantic information from visual inputs, and achieve an adapted visual representations $x^{V \rightarrow L}$ with the correct dimension.

Given $x^{V \rightarrow L}_i$ and the style vector of another factual caption $y_{i-1}$, the generator aims to reconstruct $y_i=\{d^i_{1}, \ldots, d^i_{k_i}\}$, forming the captioning loss as:
\begin{equation}
\begin{split}  
\mathcal{L}_{Cap}&=\mathcal{L}_\textrm{CE}\bigg(y_i,G\Big(E_c\big(x^{V \rightarrow L}_i\big),E_s\big(y_{i-1}\big)\Big)\bigg)\\
& = -\sum_{n=1}^{K_i} \textrm{log } p_{\theta_{M^{V \rightarrow L}},\theta_{G}, \theta_{E_c}}(d_n^i|d_{1:n-1}^i,x_i,y_{i-1}).
\end{split}
\end{equation}

To better align projected visual representations with their related text embeddings, we optimize the L2 loss between the content vectors that are extracted 
from $x^{V \rightarrow L}_i$ and the text embedding of its factual caption $y_i$. The loss function is as follows:
\begin{equation}
\begin{split}  
\mathcal{L}_{V2L} &= \Big|\Big| E_c\big(x^{V \rightarrow L}_i\big),E_c\big(y_{i}\big)	\Big|\Big|.
\end{split}
\end{equation}

Our final visual loss function is as follows: 
\begin{equation}
\begin{split}  
\mathcal{L}_{Visual} & = (\mathcal{L}_{Cap} + \mathcal{L}_{V2L})/2.
\end{split}
\end{equation}
\paragraph{\textbf{Multi-task Training Procedure}} To ensure the ability of cross-modal alignment and text style injection at the same time, we apply a multi-task training strategy. The overall loss function is as follows:
\begin{equation}
    \mathcal{L}(\theta_{M^{V \rightarrow L}},\theta_G, \theta_{E_c}) = \mathcal{L}_{Visual} + \mathcal{L}_{Text},
\end{equation}
where $\theta_{M^{V \rightarrow L}}, \theta_G, \theta_{E_c}$ are the parameters of visual projector $M^{V \rightarrow L}$, caption generator $G$ and content extractor $E_c$. 
\subsection{Inference Procedure} \label{sec:inference} Previous methods for stylized visual captioning aim to achieve several style-related knowledge for a fixed set of discrete styles. In contrast, our extracted style vector can represent multiple text style attributes simultaneously \cite{riley2020textsettr}. For example, an input sentence can cover the style of being positive, informal, etc.

During inference, FS-StyleCap accepts a few example sentences provided by users, which may represent a combination of different style attributes. We use our style extractor to extract style vectors from each example sentence and then mean pool them to obtain the target style vector, denoted as $s_{tgt}$. 
To inject a target style, we need to adjust the style vector towards the appropriate direction. 
Specifically, we move away from the factual style that describes visual content objectively and neutrally, and towards the target style to some degree.
The style vector $s$ used for decoding is computed as $\lambda (s_{tgt} - s_{src}) + s_{src}$, where $s_{src}$ is the style vector of factual style, and the delta scale $\lambda$ is a hyper-parameter. Generally, $\lambda$ is an integer in the range of $[1,10]$, which can control the degree of stylization. In our experiments, we choose the value that achieves the best overall performance. The stylized caption is generated based on $s$ and the visual content vectors extracted from $x^{V \rightarrow L}$.

\section{Experiments}
\begin{table*}[ht]
\caption{We compare our proposed \textbf{FS-StyleCap} with other methods on SentiCap. w.r.t. B-3 (BLEU-3), M (METEOR), C (CIDEr), CLIPS (CLIPScore), sACC (style accuracy), GM1 (geometric mean of sACC and CIDEr), GM2 (geometric mean of sACC and CLIPS). Full methods are trained using the entire text-only corpus with a style label; in the Few-Shot setting, CapDec is trained with the examples, while our model directly applies them to guide the generation process. The number of samples is displayed in the ``Data'' column. Methods marked with an asterisk (*) can handle multiple styles with a single model, but only our model is capable of handling arbitrary styles. We \sethlcolor{bestresult}\hl{color} the best overall scores in Few-Shot setting.}
\vspace{-8pt}
\label{table:senticap}
\fontsize{7.5}{13}\selectfont
\centering
\begin{tabular}
{cc|c|ccccccc|c|ccccccc}
\toprule
\multicolumn{2}{c|}{} & \multicolumn{8}{c|}{\textbf{Positive}} & \multicolumn{8}{c}{\textbf{Negative}} \\ \midrule
  &  & \multicolumn{1}{c|}{} & \multicolumn{4}{c}{\cellcolor[HTML]{F1F1F1}\textbf{\textit{Content}}} & \cellcolor[HTML]{F8F8F8}\textbf{\textit{Style}} & \multicolumn{2}{c|}{\cellcolor[HTML]{F1F1F1}\textbf{\textit{GEOMEAN}}} & \multicolumn{1}{c|}{} & \multicolumn{4}{c}{\cellcolor[HTML]{F1F1F1}\textbf{\textit{Content}}} & \cellcolor[HTML]{F8F8F8}\textbf{\textit{Style}} & \multicolumn{2}{c}{\cellcolor[HTML]{F1F1F1}\textbf{\textit{GEOMEAN}}} \\
\multirow{-2}{*}{} & \multirow{-2}{*}{\textbf{Model}} & \multicolumn{1}{c|}{\multirow{-2}{*}{\textbf{\begin{tabular}[c]{@{}c@{}}Data \end{tabular}}}} & \textbf{B-3} & \textbf{M} & \textbf{C} & {\textbf{CLIPS}} & \textbf{sACC} & {\textbf{GM1}} & \multicolumn{1}{c|}{{\textbf{GM2}}} & \multicolumn{1}{c|}{\multirow{-2}{*}{\textbf{\begin{tabular}[c]{@{}c@{}}Data\end{tabular}}}} & \textbf{B-3} & \textbf{M} & \textbf{C} & {\textbf{CLIPS}} & \textbf{sACC} & {\textbf{GM1}} & {\textbf{GM2}} \\

 \midrule
\multicolumn{1}{c|}{\multirow{4}{*}{\rotatebox{90}{Few-Shot}}} & CapDec \cite{nukrai2022capdec} &  100  & 16.8 & 17.4 & 52.9	& 63.9 & 79.5 & 64.9 & 71.3 & 100 	& 14.1 & 14.5 & 48.8 & 64.5 & \textbf{76.9} & 61.3 & 70.4\\
\multicolumn{1}{c|}{} & \textbf{FS-StyleCap*} &  1  & \textbf{19.2} & 17.3 & 65.3 & \textbf{73.7} & 70.8 & 68.0 & 72.2 & 1  & 20.4 & 17.6 & 71.6 & 74.3 & 64.1 & 67.7 & 69.0\\
\multicolumn{1}{c|}{} & \textbf{FS-StyleCap*} &  5 & 18.8 & \bf 18.0 & \textbf{65.6} & 73.5 & 72.3 & 68.9 & 72.9 & 5  & \textbf{20.6} & \textbf{17.9} & \textbf{71.7} & \textbf{74.6} & 67.5 & 69.6 & 71.0\\
\multicolumn{1}{c|}{} & \textbf{FS-StyleCap*} &  100  & 17.7 & 17.9 & 62.8 & 72.7 & \textbf{82.8} & \textbf{\sethlcolor{bestresult}\hl{72.1}} & \textbf{\sethlcolor{bestresult}\hl{77.6}} & 100  & 20.0 & 17.3  & 68.8 & 74.4 &75.4  & \textbf{\sethlcolor{bestresult}\hl{72.0}} & \textbf{\sethlcolor{bestresult}\hl{74.9}} \\
\midrule
\midrule
\multicolumn{1}{c|}{\multirow{3}{*}{\rotatebox{90}{Full}}} & StyleNet \cite{gan2017stylenet} &  2,994   & 12.1 & 12.1 & 36.3 & - & 45.2 & 40.5 &- & 2,991  & 10.6 & 10.9 & 36.6 & - & 56.6 & 45.5 & - \\
\multicolumn{1}{c|}{} & MSCap* \cite{guo2019mscap} &  2,994 & 16.2 & 16.8 & 55.3 & - & \textbf{92.5} & 71.5 & - & 2,991  & 15.4 & 16.2 & 51.6 & - & \textbf{93.4} & 69.4 &- \\
\multicolumn{1}{c|}{} & CapDec \cite{nukrai2022capdec} &  2,994  & \textbf{26.9} & \textbf{20.0} & \textbf{64.5} & \textbf{72.4 }& 87.0 & \textbf{74.9} & \textbf{79.4} & 2,991  & \textbf{19.9}	& \textbf{19.0} &	\textbf{65.8} &	\textbf{71.5} &	79.7 & \textbf{72.4} & 	\textbf{75.5} \\
\bottomrule
\end{tabular}

\end{table*}

\subsection{Datasets}
\noindent\textbf{\textit{MSCOCO}} \cite{lin2014microsoft}. For image captioning task, we leverage the popular MSCOCO datatset. We apply the Karpathy splits \cite{karpathy2015deep} which contains 113,287 images for training, 5,000 images for validation, and  5,000 images for testing. Each image contains 5 captions. 

\noindent\textbf{\textit{MSR-VTT}} \cite{xu2016msr}. For video captioning task, we leverage the MSR-VTT dataset, which includes 6,513 videos for training, 497 videos for validation, and 2,990 for testing. Each video contains 20 captions. 

\noindent\textbf{\textit{Personality-Captions}} \cite{shuster2019engaging}. Our unlabeled textual training corpus comes from the Personality-Captions dataset, which contains sentences conditioned on possible personality traits. We also add 1,000 factual descriptions, enabling our style extractor to handle factual style. During training, we remove the style labels and only use pairs of adjacent sentences from each personality set as input samples. We pre-process all texts to meet the normalized format \cite{li2018delete}, resulting in 136,894 training and 3,683 validation examples.

\noindent\textbf{\textit{SentiCap}} \cite{mathews2016senticap}. To evaluate the results of stylized image captioning, we apply the test set of SentiCap, a sentimental image captioning dataset developed from the MSCOCO dataset. Each image in the dataset is labeled with three positive and three negative captions. The positive and negative subsets have 998/673 and 997/503 images for training/testing, respectively.

\subsection{Implementation Details}
In our image captioning experiments, we extract image features using the CLIP (ViT-B/32) model \cite{radford2021clip}. For video captioning experiments, we follow \citet{tang2021clip4caption}, using TSN sampling to sample at most 20 frames from the video. We then encode these frames using a pre-trained CLIP (ViT-B/32) video encoder \cite{luo2022clip4clip}, and concatenate the outputs to form our visual input. During training, we apply T5-base as our pre-trained language model, which is fine-tuned for 20 epochs, with a batch size of 128. The learning rate for visual projection module is set to $1e-3$, while the others are set to $5e-4$. During inference, we sample 100 positive/negative examples from the training set of SentiCap to guide the stylized generation. We conduct experiments for multiple styles using 1 and 5 examples respectively. The examples are presented in our supplementary material.

\subsection{Baseline Models}
Stylized Image Captioning baseline models for comparison include:

\noindent\textbf{\textit{StyleNet}} \cite{gan2017stylenet}. This model proposes to learn different groups of matrices to capture factual and stylized knowledge.

\noindent\textbf{\textit{MSCap}} \cite{guo2019mscap}. This method is able to handle different predefined styles with a single model. They propose an adversarial network to improve the overall performances. 

\noindent\textbf{\textit{CapDec}} \cite{nukrai2022capdec}. This model applies the image features encoded by a pre-trained CLIP model and fine-tunes their text decoder from GPT-2 (large) \cite{solaiman2019release}, which achieves state-of-the-art performance on stylized image captioning. We compare our results with their model which is fully trained on the stylized corpora, as well as their model trained with only a few sentences.

For Stylized Video Captioning, the Sentimental Transformer \cite{wu2023sentimental} is used as the baseline, which integrates information from multiple modalities and incorporates prior sentimental knowledge to generate sentimental video descriptions. For fair comparison, we compare our model's results with their results when the audio modality is not used.

\begin{table}[t]
\caption{Automatic evaluation results on MSCOCO. The Factual method is specifically trained for factual image captioning, while the Stylized methods are capable of generating factual or stylized captions, and the best results are bolded.}
\vspace{-8pt}
\label{table:mscoco}
\centering
\fontsize{7.5}{11}\selectfont
\begin{tabular}{c|c|cccc}
\toprule
 & \textbf{Model} & \textbf{BLEU-4} & \textbf{METEOR} & \textbf{CIDEr} & \textbf{SPICE} \\
 \midrule
Factual & ClipCap \cite{mokady2021clipcap} & 33.53 & 27.45 & 113.08 & 21.05 \\
\midrule
\multirow{4}{*}{Stylized} & StyleNet \cite{gan2017stylenet} & 21.18 & 20.47 & 66.42 & 13.54\\
 & SemStyle \cite{mathews2018semstyle} & 23.79 & 21.91 & 76.97 & 15.75\\
 & DetachAttach \cite{tan2022detach} &  20.43 & 21.84 &  76.91 & 16.21\\
 & \textbf{FS-StyleCap} & \textbf{30.38} & \textbf{26.76} & \textbf{105.18} & \textbf{19.00}\\
\bottomrule 
\end{tabular}
\vspace{-3pt}
\end{table}

\subsection{Evaluation Metrics}

Our \textbf{automatic evaluation metrics} for Few-Shot SVC consider two aspects:
the ability to generate fluent and relevant descriptions for visual content, and the ability to express in a target style.

Following previous methods \cite{guo2019mscap,li2021similar,tan2022detach}, we evaluate the sentence relevancy and fluency using widely used metrics \footnote{As in previous works, we utilize the open-source evaluation code from: https://github.com/jmhessel/pycocoevalcap.}, 
 including BLEU \cite{papineni2002bleu}, METEOR \cite{banerjee2005meteor}, and CIDEr \cite{vedantam2015cider}. We also report the perplexity scores in the supplementary material, which are evaluated by SRILM \cite{stolcke2002srilm}. In addition, we compute the CLIPScore \cite{hessel2021clipscore} to show the visual relevance, which measures the similarity between an image and its stylized caption.
To evaluate the style accuracy (sACC), we apply a pre-trained style classifier following previous works \cite{tan2022detach,riley2020textsettr}. This classifier is fine-tuned from a BERT-base model \cite{devlin2018bert}, using the training set of SentiCap. It achieves an accuracy of $98\%$ on the validation set.
Moreover, we follow previous works \cite{tan2022detach,hu2022text} to evaluate  the \textbf{overall performance} using the geometric mean score of semantic relevance and style accuracy, i.e. the geometric mean score of sACC and CIDEr.

In order to measure the effectiveness of our model to handle multiple styles, we also perform \textbf{human evaluation} in term of two aspects: (1) Visual Relevance that measures relationship between the stylized captions and source image. (2) Style appropriateness which means how well the descriptions express the target style specified by the example sentences. Following the standard in \cite{guo2019mscap,tan2022detach}, relevance is rated from 0 (unrelated) to 3 (very related), and style appropriateness from 0 (bad) to 3 (perfect). We conduct experiments on three models with six styles, and sample $5\%$ of the images from the test set of SentiCap. This results in $30 \times 6 \times 3$ Image-Stylized Caption pairs for human evaluation. 

\section{Results and Discussion}

\subsection{Automatic Metric Results}
To evaluate the performance of our model in generating stylized visual captions, we conduct experiments on both image captioning and video captioning. 

Specifically, for sentimental image captioning, where a human-annotated dataset exists, we compare our results with other baselines on SentiCap. The full methods are trained on the entire text-only training set from SentiCap, while the few-shot methods only use a small number of stylized samples from the same set. Moreover, our proposed \textbf{FS-StyleCap} can handle multiple styles using a single model and do not require additional training for few-shot inference. As illustrated in Table \ref{table:senticap}, FS-StyleCap outperforms CapDec which is trained with the same $100$ stylized examples, both in visual relevance and style accuracy. We also surpass other full methods MSCap and StyleNet, and achieve a comparable overall performance to the state-of-the-art method CapDec when it is fully trained with a labeled corpus.  To measure our ability to extract text style from even fewer number of examples, we conduct experiments using only 1 and 5 example sentences, respectively. These experiments show credible results as well, better than CapDec trained with 100 samples.
In addition, as \textit{factual} can also be figured as a style \cite{tan2022detach,mathews2018semstyle}, 
we generate factual captions and evaluate their semantic relevance on MSCOCO. As shown in Table \ref{table:mscoco}, our model outperforms previous methods that can handle factual and other styles, and is comparable to methods that are specifically trained for standard image captioning, such as ClipCap \cite{mokady2021clipcap}.

\begin{table}[t]
\caption{Evaluation results on MSR-VTT. We apply the factual ground truths to evaluate the semantic
relevance.}
\vspace{-8pt}
\label{table:msrvtt} 
\centering
\fontsize{7.5}{10}\selectfont
\begin{tabular}{ccccccc}
\toprule
\multicolumn{7}{c}{\textbf{Positive}} \\
\midrule
\multicolumn{1}{c|}{\textbf{}} & \multicolumn{1}{c|}{\textbf{Model}} & \multicolumn{1}{c|}{\textbf{Data}} & \textbf{B-3} & \textbf{C} & \textbf{sACC} & \textbf{GM1} \\
\midrule
\multicolumn{1}{c|}{\multirow{3}{*}{Few-Shot}} & \multicolumn{1}{c|}{\textbf{FS-StyleCap}} & \multicolumn{1}{c|}{1} & \textbf{38.0} &	33.8 &	58.9 &	44.6 \\
\multicolumn{1}{c|}{} & \multicolumn{1}{c|}{\textbf{FS-StyleCap}} & \multicolumn{1}{c|}{5} & 37.2 & 32.7 &	61.0 &	44.7 \\
\multicolumn{1}{c|}{} & \multicolumn{1}{c|}{\textbf{FS-StyleCap}} & \multicolumn{1}{c|}{100} & 32.0 &	\textbf{36.6}	&72.8&	\textbf{\sethlcolor{bestresult}\hl{51.6}} \\
\midrule
\multicolumn{1}{c|}{Full} & \multicolumn{1}{c|}{Senti-Trans \cite{wu2023sentimental}} & \multicolumn{1}{c|}{2,994} & 28.9 & 25.0 & \textbf{97.8} & 49.5 \\
\midrule
\multicolumn{7}{c}{\textbf{Negative}} \\
\midrule
\multicolumn{1}{c|}{\multirow{3}{*}{Few-Shot}} & \multicolumn{1}{c|}{\textbf{FS-StyleCap}} & \multicolumn{1}{c|}{1} & \textbf{48.2}	& 43.6	&65.5 &	53.4 \\
\multicolumn{1}{c|}{} & \multicolumn{1}{c|}{\textbf{FS-StyleCap}} & \multicolumn{1}{c|}{5} & 47.9 &	45.0	& 71.8 &	56.8\\
\multicolumn{1}{c|}{} & \multicolumn{1}{c|}{\textbf{FS-StyleCap}} & \multicolumn{1}{c|}{100} & 42.0 &\textbf{46.1}	&72.7	&\textbf{\sethlcolor{bestresult}\hl{57.9}} \\
\midrule
\multicolumn{1}{c|}{Full} & \multicolumn{1}{c|}{Senti-Trans \cite{wu2023sentimental}} & \multicolumn{1}{c|}{2,991} & 25.9 & 16.7 & \textbf{96.2} & 40.08 \\
\bottomrule
\end{tabular}
\vspace{-4pt}
\end{table}

For video captioning, since there is no dataset with paired sentimental captions, we use the factual ground truths in MSR-VTT to evaluate the visual relevance. As illustrated in Table \ref{table:msrvtt}, our proposed FS-StyleCap outperforms the full method Senti-Transformer, especially for the negative style. We find that a small set of captions in MSR-VTT tend to represent positive styles, such as ``pretty girl'', ``nice car'', etc. Our inference method (Section \ref{sec:inference}) can distinguish these styles to some extent by subtracting the style vector of MSR-VTT, while Senti-Transformer might not be able to do this.


\subsection{Human Evaluation Results}\label{sec:human_eval}
We conduct human evaluation experiments on six styles: (1) two standard styles of positive and negative; (2) four more fine-grained styles of romantic, adventurous, memorable and skeptical. For each style, 1-5 example sentences are provided as examples (as shown in Figure \ref{fig:task}). We compare our results with CapDec, trained with 100 example sentences. We also compare our model with a powerful large language model (LLM). We use ClipCap to generate factual captions for images, and then apply Copilot Hub \footnote{https://app.copilothub.ai/home} to perform text style transfer with 100 examples. Specially, we upload the stylized examples to train the model to incorporate the target style. This model is based on ChatGPT, specially designed for stylized generation. Following previous methods \cite{suzgun2022prompt,reif2021recipe}, we design several prompts for text style transfer, and finally use the following one which achieves the best performance: 
\begin{itemize} [itemsep=0.1em,parsep=0em,topsep=0em,partopsep=0em,leftmargin=1em,itemindent=0.2em]
    \item Here is a text, which is a COCO image caption: "\{input\}". Here is a rewrite of the text, which is more in your style but keeps its semantics: "
\end{itemize} The human evaluation results are obtained from 13 English-fluent volunteers. As shown in Figure \ref{fig:human_evaluation}, our method achieves the best overall performance, particularly in fine-grained styles. Our method outperforms other methods in terms of style accuracy, and its visual relevance is comparable to the results based on LLM. We find that training with numerous examples can improve the performance of the large language model and CapDec beyond their current capabilities.
However, this requires users to curate a larger number of style-labeled examples (i.e. from 5 to 2,000) and wait for additional training, which is less flexible.


\subsection{Ablation Study}
We carry out several ablation studies to show the effectiveness of our proposed components:
\begin{figure}[t]
    \centering
\includegraphics[width=0.9\linewidth]{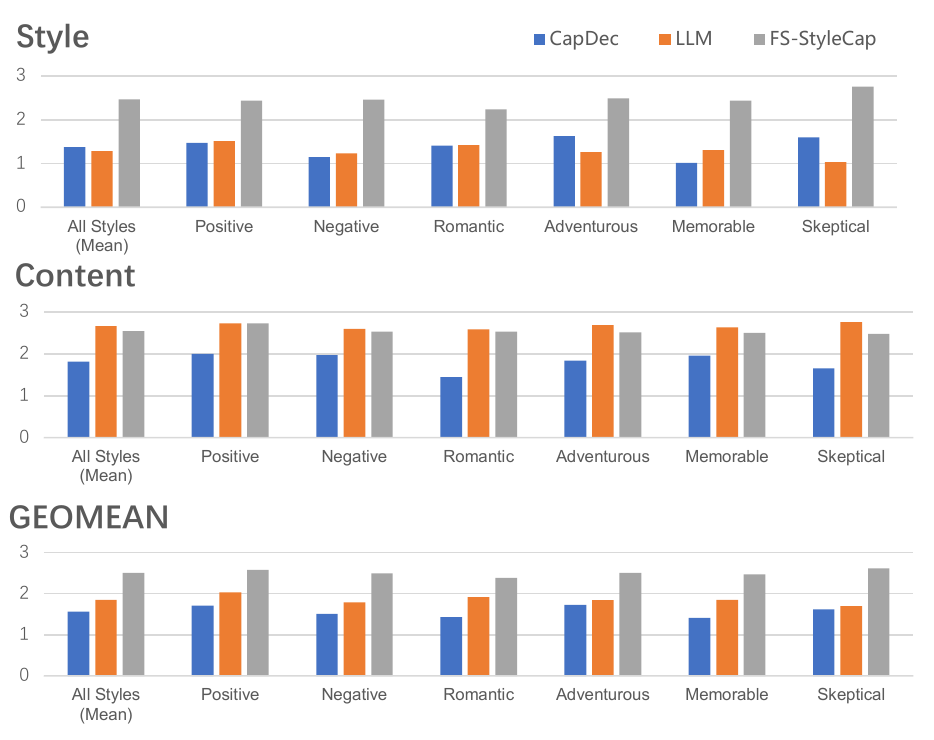}
\vspace{-8pt}
    \caption{Human evaluation results for six styles. GEOMEAN is measured by the geometric mean of Style (style appropriateness) and Content (visual relevance).}
    \label{fig:human_evaluation}
\end{figure}

\begin{table}[t]
\centering
\fontsize{8}{12}\selectfont
\caption{\label{ablation_table}
Ablation studies on positive and negative style. The overall metric GM1 is measured by the geometric mean of CIDEr and sACC.}
\vspace{-8pt}
\begin{tabular}{l|ccccc}
\toprule
\textbf{Model} &\textbf{CIDEr$\uparrow$} & \textbf{sACC$\uparrow$} &\textbf{GM1$\uparrow$} \\
\midrule
\textbf{FS-StyleCap} & 66.26 &	79.10 & \textbf{72.40}\\
\midrule
w/o $\mathcal{L}_{DR}$ & 38.60	& \textbf{87.30} &	58.05\\
w/o $\mathcal{L}_{NBT}$  & 67.05	& 	67.85 &	67.45\\
w/o $\mathcal{L}_{Style}$ & \textbf{67.06} &	69.80 &	68.41\\
w/o $\mathcal{L}_{V2L}$ & 54.75 &	75.70 &	64.38\\
w/o $MultiTask$ & 2.45& 	81.20	& 14.10\\
\bottomrule 
\end{tabular}
\end{table}

\begin{figure}[h]
    \centering
\includegraphics[width=0.7\linewidth]{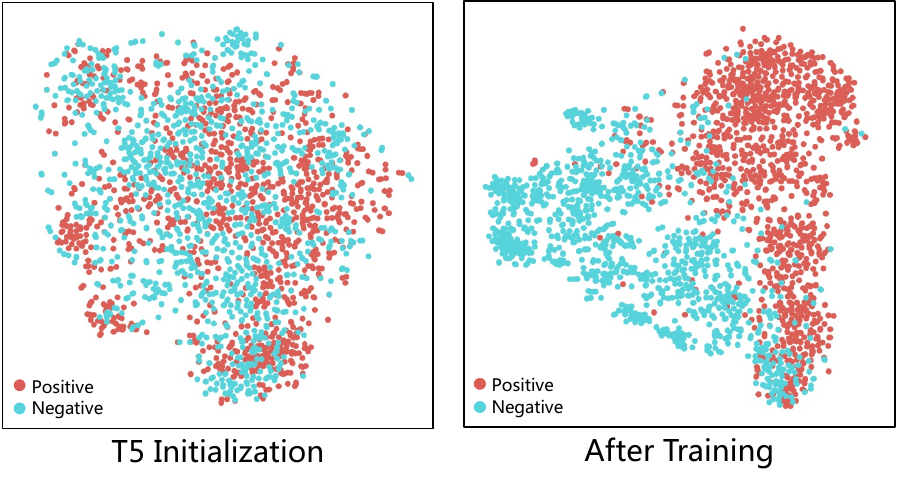}
\vspace{-12pt}
    \caption{tSNE plots for the embeddings of the extracted style vectors of positive and negative samples.}
    \label{fig:embedding}
\vspace{-4pt}
\end{figure}

\begin{figure*}[ht]
    \centering
    \includegraphics[width=0.9\textwidth]{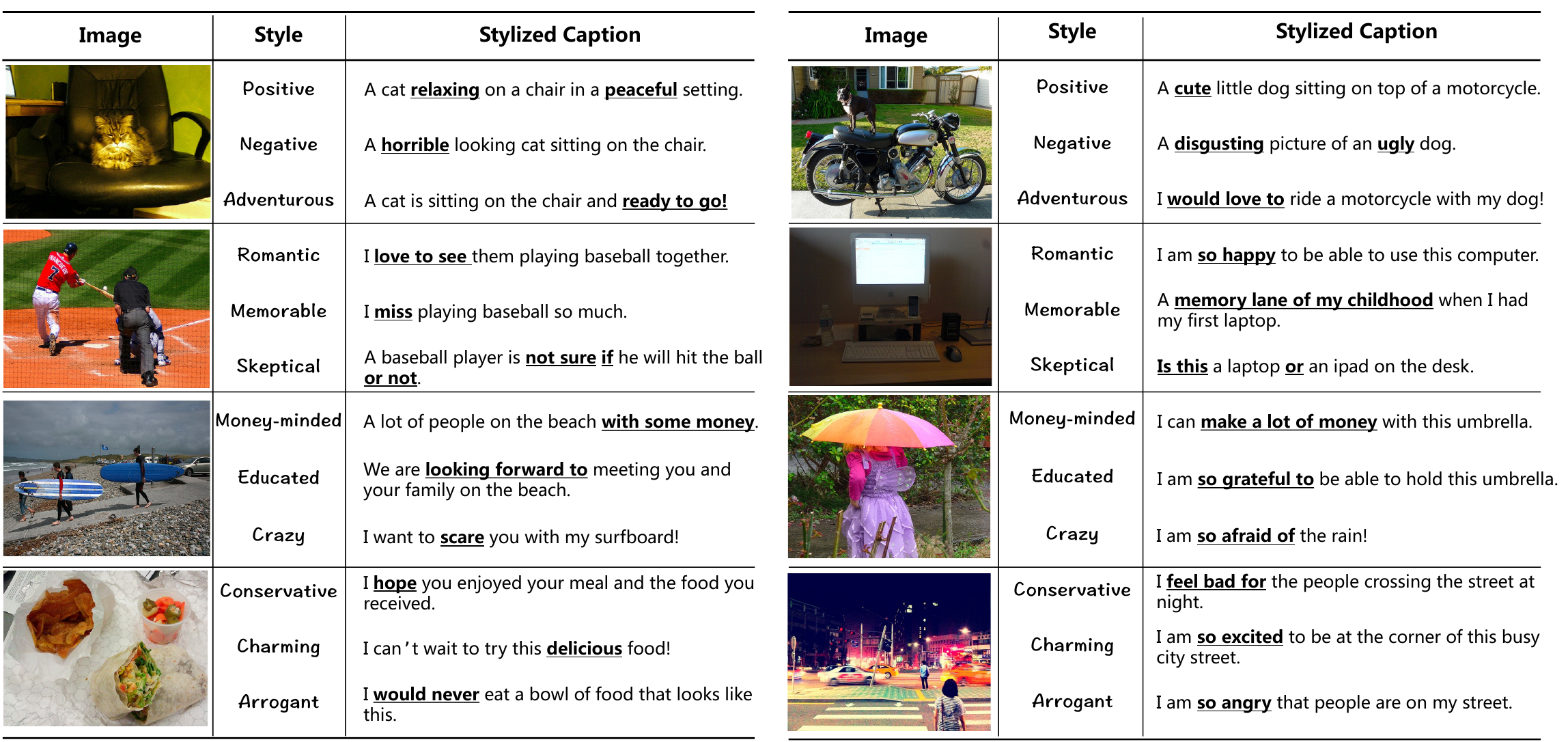}
    \vspace{-7pt}
    \caption{Examples of stylized captions generated by our proposed FS-StyleCap. Each style is guided by 5 examples, which are provided in our supplementary material.}
    \label{fig:img_cases}
\end{figure*}

\begin{figure*}[ht]
    \centering
    \includegraphics[width=0.92\textwidth]{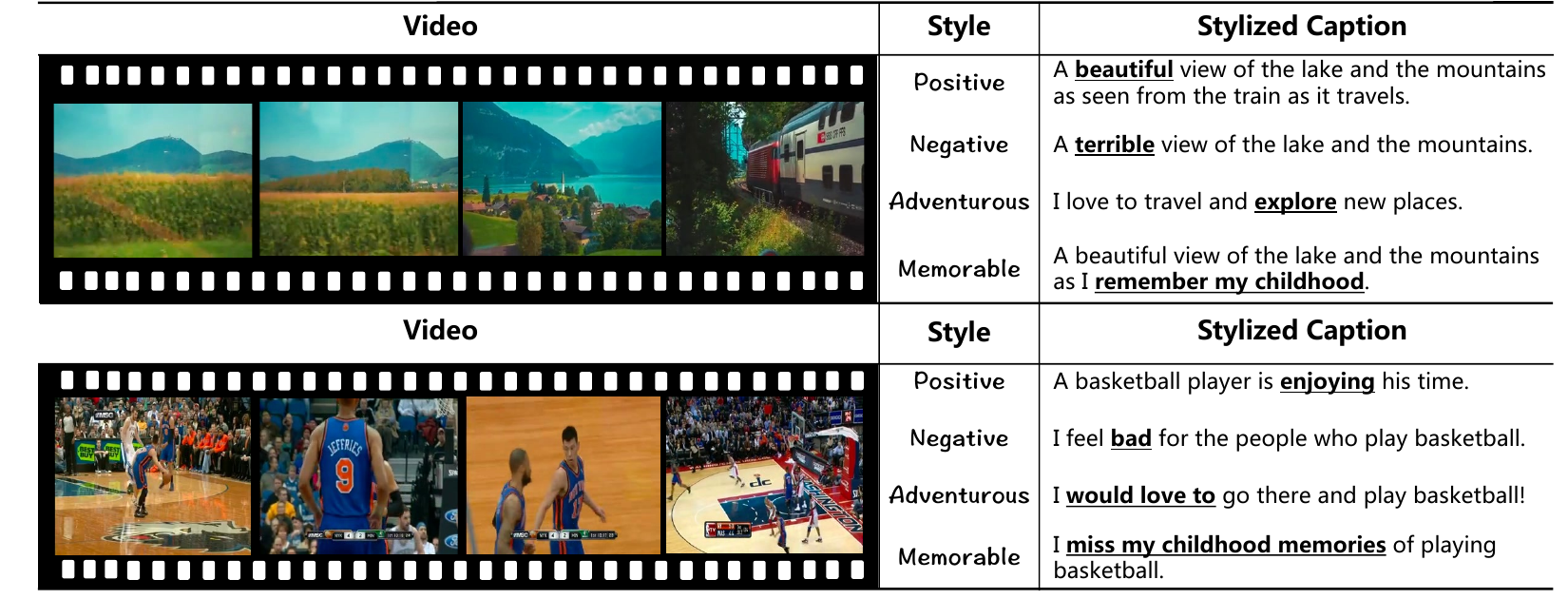}
    \vspace{-8pt}
    \caption{Illustration of Stylized Video Captioning examples by our FS-StyleCap.}
    \label{fig:video_cases}
    \vspace{-3pt}
\end{figure*}

\parskip=0.1em
\begin{itemize} [itemsep=0.1em,parsep=0em,topsep=0em,partopsep=0em,leftmargin=1em,itemindent=0.2em]
    \item \textbf{w/o $\mathcal{L}_{DR}/\mathcal{L}_{NBT}/\mathcal{L}_{Style}$.} To evaluate the effectiveness of each loss function for text style injection, we remove the denoising reconstruction loss ($\mathcal{L}_{DR}$), the noisy back translation loss ($\mathcal{L}_{NBT}$) and the style discrimination loss ($\mathcal{L}_{Style}$) from the objective function (Eqn \ref{eqn:l_text}).
    \item \textbf{w/o $\mathcal{L}_{V2L}$.} To improve the visual relevance, we add the V2L loss, which further aligns the projected visual embeddings to its correlated text embeddings. In order to show its efficiency, we remove this loss and compare the experimental results.
    \item \textbf{w/o MultiTask} To verify the effectiveness of the multi-task training strategy in our second training stage, we freeze all of the language models trained in stage 1, and only trains the visual projection module with the task of factual visual captioning. 
\end{itemize}
Table \ref{ablation_table} presents our ablation studies on the stylized image captioning task. We observe that: (1) Noisy back translation helps improve style injection quality, while noisy reconstruction helps maintain semantic relevance. With the combination of them, we could achieve a better result; (2) Style discrimination loss significantly improves the style accuracy; (3) Removing $\mathcal{L}_{V2L}$ reduces the visual relevance, which demonstrates its effectiveness to make cross-modal alignment; (4) As our model's parameters and the pre-training data in stage one are not so large, if we freeze the parameters of our language model ($E_s,E_c,G$) and only update the parameters of $M^{V \rightarrow L}$, the performance degrades significantly. In summary, our designed components/schemes all contribute to our FS-StyleCap performance.

\subsection{Embedding Visualization}
To demonstrate the effectiveness of our extracted style vectors, we sample $1,000$ positive examples and $1,000$ negative examples, and visualize the distributions of their style vectors using tSNE plots~\cite{van2008visualizing}. The results shown in Figure \ref{fig:embedding} display that after fine-tuning with our proposed methods (refer to Section \ref{sec:style_extraction}), the style extractor is capable of separating different style attributes in the embedding space, which demonstrates its effectiveness in extracting information related to style.

\subsection{Qualitative Results}
Figure \ref{fig:img_cases} presents some stylized image captioning cases. We illustrate several styles to show that our model can handle multiple styles with a single model, and does not require further training for new style. 
Figure \ref{fig:video_cases} shows some cases of stylized video captioning. More cases can be found in our supplementary material.

\section{Conclusion and Future Work}
In this paper, we tackle the task of Few-Shot Stylized Visual Captioning, which aims to describe images and videos with the text style guided by a few examples. We propose a framework FS-StyleCap, which consists of a conditional encoder-decoder language model and a visual projection module. We firstly train the style extractor on an unlabeled text corpus with proposed losses. After achieving the extractor to extract style representation, we address the problem of cross-modal alignment, enabling our model to generate stylized descriptions guided by any desired style. Experiments on automatic evaluation and human evaluation demonstrate the effectiveness of our method in generating captions with multiple desired styles, only guided by a few examples. In future work, we intend to experiment with larger text corpora and larger pre-trained models to expand the capabilities of FS-StyleCap, allowing it to handle more challenging styles. Additionally, we believe that our method has the potential to support multi-modal inputs and we plan to explore this possibility as well.

\bibliographystyle{ACM-Reference-Format}
\bibliography{refs}
\end{document}